# Deformation potential dominated phonons in ZnS quantum dots

S. Dhara,[1, 2,*] A. K. Arora,[2] Jay Ghatak,[3] K. H. Chen,[4, 5] C. P. Liu,[6] L. C. Chen,[5] Y. Tzeng[1] and Baldev Raj[2]

[1] Department of Electrical Engineering, Institute for Innovations and Advanced Studies, National Cheng Kung University, Tainan-701, Taiwan.

[2] Materials Science Division, Indira Gandhi Centre for Atomic Research, Kalpakkam 603 102, India

[3] Institute of Physics, Sachivalaya Marg, Bhubeneswar, India

[4] Institute of Atomic and Molecular Sciences, Academia Sinica, Taipei-106, Taiwan

[5] Center for Condensed Matter Sciences, National Taiwan University, Taipei-106, Taiwan

[6] Department of Materials Science and Engineering, National Cheng Kung University, Tainan-701, Taiwan.

## *Abstract*

Strong deformation potential (DP) dominated Raman spectra are reported for quantum confined cubic ZnS nanoclusters under off-resonance conditions allowed only in quantum dots. A flurry of zone boundary phonons is demonstrated in the scattering process. Transverse optic (TO) mode in the multi-phonon process shows only even order overtones suggesting the dominance of a two-phonon process (having large DP value in ZnS) and its integral multiples. Two-phonon TO modes corresponding to $A_1$ and $B_2$ symmetries are also demonstrated under off-resonance conditions which are allowed only in quantum dots.

---

[*] Communicating author email: dr.s.dhara@gmail.com

Quantum physics and its applications, particularly for zero-dimensional (0-D) quantum dots (QDs) with radius close to or less than the Bohr excitonic radius, are interesting [1]. While there are plenty of studies available for confinement effects of electronic states, interest in phonon confinement is on the surge. Phonon confinement effect results in the breakdown of q = 0 selection rule owing to finite crystallite size [2]. Observations related to the broadening and red shift of phonon frequency [2], and the appearance of zone boundary (ZB) phonons [3] or acoustic phonons as combination modes [4] are well documented. Additional peaks close to $1^{st}$ order optical Raman modes are also reported as surface optic (Frohlich) modes [5] and exciton-phonon (polaron) coupled modes [6]. The breakdown of selection rules close to resonance in the Raman scattering study of quantum well (2-D) structures is reported in case of GaAs/AlAs superlattice structures [7] for both the longitudinal optical (LO) and transverse optical (TO) modes. In cubic crystals of zinc-blend structure, the Raman phonons have $F_2$ modes ($T_d$ point group) in the confined configurations of even ($A_1$) and odd ($B_2$) order symmetries. In 2-D electronic confinement, only even orders corresponding to $A_1$ symmetries appear in both polarization modes. The deformation potential (DP, a short-range interaction between the lattice displacement and the electrons) also plays an important role at resonance in all the TO modes corresponding to both $A_1$ and $B_2$ symmetries, which are observable under polarized conditions for the confined superlattice structure [7]. In 2-D confinement, all the above observations are made with finite value of wave vector $k_z$ (restricted by few repeating layers in the superlattice structure) and not too large values of $k_{x,y}$. Thus, in the 0-D configuration (QDs) with limited number of unit cells (i.e. finite values of $k_{x,y,z}$ vectors) the matrix element of the electron-phonon interaction becomes truly $k$-independent and can also be represented only by a DP [8]. The resulting Raman resonance can be very strong with only the $A_1$ symmetries observable in the unpolarized configuration while both $A_1$ and $B_2$ symmetries are present only under off-resonance conditions.

We examine the validity of selection rules in 0-D systems away from resonance, as discussed above, for Raman spectra of ZnS quantum dots. In order to understand scattering mechanism related to DP at reduced dimension, ZnS is chosen for having large DP value at least in the two-phonon process [9]. The band-gap of bulk cubic ZnS is ~337 nm (3.68 eV) and the Bohr exciton radius is ~1.25 nm [10]. We observe a strong multi-phonon scattering of TO modes in the quantum confined nanoclusters of cubic ZnS (diameter ~ 2.2 nm having no more than 12 unit cells) far from the resonance conditions. We observe, for the first time, the dominant roles of DP (overtones in TO modes) and the presence of TO phonon modes belonging to both $A_1$ and $B_2$ symmetries in the 0-D system under off-resonance condition.

ZnS QDs, used in the present study, are grown as free standing powder by a wet chemical synthesis method using $ZnCl_2$ as the precursor and $Na_2S$ as a source of sulfur [11]. Triethylamine (TEA) is used as capping agent for controlling the cluster size. Besides, Zinc acetate, $Zn(Ac)_2$ as precursor and thioacetamide (TAA) as capping agent are also used for microwave based synthesis of ZnS QDs of diameter of ~ 2.2 nm [12].

Structural studies using x-ray diffraction (XRD; STOE Diffractometer) show diffraction peaks at $2\theta$ values of $28.80^o$, $48.5^o$, and $56.75^o$ (Fig.1a). These peak positions correspond to reflections from crystalline planes (111), (220) and (311), respectively, for the cubic phase of ZnS [12]. Broadening of the peaks (Fig. 1a) indicates the nanocrystalline nature of the sample. The average crystallite sizes are calculated from the full width at half maximum (FWHM) of the diffraction peaks using the Scherrer formula, $D = k\lambda/\beta\cos\theta$, where $D$ is the mean particle diameter, $k$ is a geometric factor (equal to 0.89), $\lambda$ is the X-ray wavelength, $\beta$ is the FWHM of diffraction peak, and $\theta$ is the diffraction angle. The typical grain sizes of the ZnS particles calculated from the most intense peaks are approximately 2.2-5.0 nm. Transmission electron microscopic (TEM; JOEL 2000FX) study shows (Fig. 1b) typical clusters of $D \sim 2.2$ nm. Continuous rings in the selected area electron diffraction (SAED) pattern from small clusters corresponding to crystalline planes (111), (200), (311) and (400) of cubic ZnS phase are shown in the outset (Fig. 1b). High resolution TEM (HRTEM; JEOL TEM 2010 UHR)  analysis of clusters with $D \sim 2.5$ nm size shows $d$-spacing of ~0.31 nm and ~0.268 nm corresponding to (111) and (200) planes of   ZnS, respectively (Fig. 1c) [13].

For the understanding of different excitation processes, we also measure the band edge shift with cluster size [14]. The typical sizes of the QDs are confirmed by analyzing ultra violet-visible (UV-Vis) absorption spectrum (Fig. 2) where the band-edge absorption shows a strong blue shift with respect to the bulk value of 337 nm [9].   Shoulders around 253 - 300 nm show the onset of UV absorption. Average sizes of the QDs are calculated to be around 2.2 -5.0 nm using tight binding approximation of effective masses ($m_e^*$ for electron   and $m_h^*$ for hole) in the expression for band edges [14], $E(D) = [2\hbar^2\pi^2(1/m_e^*+1/m_h^*)/D^2 - 3.572e^2/\varepsilon D - 0.124e^4/\hbar^2\varepsilon(1/m_e^*+1/m_h^*)]$ where $\hbar$ is the Planck's constant and $\varepsilon$ (= 8.9 for cubic ZnS) is the dielectric constant. These values of cluster sizes are in good agreement with what are calculated based on XRD analysis and TEM studies.

Higher order Raman scattering (Jobin-Yvon T64000 spectrograph) studies using 325 nm excitation for ZnS nanoclusters with different sizes show (Fig. 3) 1st order 1-TO (~285 cm$^{-1}$) and 1-LO (346 cm$^{-1}$) modes. 1-LO mode corresponds to zone centre ($\Gamma$ point of Brillouin zone) phonon. However, 1-TO mode at 285 cm$^{-1}$ originates from the ZB, as the same value is reported at $L$ point of Brillouin zone for bulk ZnS [15]. A combination mode around 420 cm$^{-1}$ [16,17] comprising of optical mode at zone center plus acoustic mode at ZB of either 1-(LO$_I$+TA$_L$) or 1-(TO$_I$+LA$_L$) are also observed to diminish with decreasing size and vanish below the confinement size along with the 1-TO mode. An excessive presence of the ZB phonons in the confined ZnS nanoclusters is further confirmed using a visible excitation. A typical 1st order Raman scattering (Jobin-Yvon LabRam HR) study with 532 nm excitation shows (inset Fig. 3) a new peak ~ 257 cm$^{-1}$ along with the 1-LO mode for the clusters below the confinement size. A combination mode of 1-(LO$_I$-TA$_L$) can explain the presence of the new peak [16]. Overall absence of 1-TO mode with visible excitation is due to poor scattering efficiency as well as anti-resonant behavior reported for the mode in cubic ZnS [16]. A systematic evolution of higher order 2-TO mode and at the same time gradual decrease of 2-LO mode with decreasing size below quantum confinement size (~2.5 nm) are also observed for the UV excitation. We may restate here that with reducing cluster size the band gap shifts to lower wavelengths (Fig. 2) i.e. further away from the resonance condition. Thus, decrease in the intensity of 2-LO modes with decreasing size (Fig. 3) can be understood under off-resonance conditions. Low value of DP, reported for 1-TO mode of cubic ZnS (25 eV [9]) may be correlated to the disappearance of 1-TO mode under off-resonance conditions for clusters below the confinement size. On the other hand, a strong DP for two-phonon process (2470 eV [9]) of cubic ZnS (two orders of magnitude higher than that of single-phonon process) along with the $k$-independence of phonon dispersion for 0-D system [8], are made responsible for the increase in the intensity of 2-TO mode (∼ 573 cm⁻¹) even under off-resonance conditions. It may be specifically noted that the 2-TO mode for the QDs with the smallest size (∼2.2 nm) shows an even higher intensity than that for the 1-LO mode (346 cm⁻¹), and 1- TO mode disappears completely along with the combination mode.   This phenomenon is analogous to the behavior of LO mode with respect to that of TO mode under resonance condition [18].

We also study the multi-phonon process, before the intensity diminishes substantially, upto 6[th] order for these clusters. Overtones of LO mode die out beyond the 2[nd] order under off-resonance conditions.   The presence of overtones corresponding to the TO modes is observed in the spectra for the clusters below the confinement size. A unusually strong presence of DP is again observed with the presence of even order overtones (Fig. 4) corresponding to the TO modes (4- TO at ∼ 1141 cm⁻¹ and 6-TO at ∼ 1720 cm⁻¹). Although these wave numbers are pretty close to fundamental frequencies and overtones (both even and odd) of LO modes reported for ZnO nanoclusters [19], a very well characterized electron energy loss spectroscopy (not shown in figure) using TEM fails to detect any signature of oxygen in our sample. However, presence of the overtones corresponding to the fundamental modes in the Raman spectrum is governed by the selection rules. In order to ascertain which   overtones of $F_2$ mode are allowed by the point group symmetry, we have determined the irreducible representations corresponding to 2-, 3-, and 4-TO modes from group theoretical analyses. The 9, 27 and 81 dimensional reducible representations associated with 2-, 3- and 4-TO modes can be written as a linear combination of the irreducible representations of the $T_d$ point group as

$F_2 \times F_2$(2-TO) = $A_1 + E + F_1 + F_2$            ….. (1)

$F_2 \times F_2 \times F_2$ (3-TO) $= A_1 + A_2 + 2E + 3F_1 + 4F_2$             ….. (2)

$F_2 \times F_2 \times F_2 \times F_2$ (4-TO) $= 4A_1 + 3A_2 + 7E + 10F_1 + 10F_2$      ….. (3)

One can see that all the overtones have Raman-active irreducible representations $A_1$, $E$ and $F_2$, suggesting that all the overtones are allowed by the selection rules. On the other hand, experiments show the existence of only even-overtones in the Raman spectra of the QDs with size below the quantum confinement. The prominent presence of even TO modes and the conspicuous absence of odd overtones under a completely off-resonance condition for QDs remind us the strong influence of DP in the multi-phonon process. Overtone is available only for 2-TO (multiples of 1, 2,.3, ..) but not for 1-TO, which has a weak DP value [9]. Under off-resonance conditions, detailed study of the 2-TO mode also shows [inset in Fig. 4] all the even- and odd-order peaks corresponding to both $A_1$ and $B_2$ symmetries, respectively, as suggested for the 0-D system [8].

Thus, we report an unprecedented role of DP as predicted by the selection rule for the quantum confined 0-D system away from the resonance. This is made possible due to excellent Raman scattering efficiency by using short wavelength photons along with the extreme case of quantum confinement effects in ZnS system having large DP in the two-phonon process. Strong Raman activities away from the resonance show promises for using QDs as analogue for the resonant Raman scattering effect, which is largely dependent on available laser excitation.

We acknowledge S. F. Chen of Department of Materials Science and Engineering, NCKU for EELS measurements and S. V. S. Rao of Centralised Waste Management Facility, Bhaba Atomic Research Centre Facilities, Kalpakkam for UV absorption measurements.

**Figure Captions :**

Fig 1. (Color online) Structural and microstructural studies of grown ZnS nanoclusters. a) X-ray diffraction pattern of cubic ZnS nanoclusters with sizes indicated in the inset. b) TEM image of nanoclusters ~2.2 nm along with (outset) a continuous ring diffraction pattern corresponding to cubic ZnS. c) HRTEM image of ZnS nanoclusters ~2.5 nm with lattice spacing of cubic ZnS.

Fig 2. (Color online) UV absorption spectra for ZnS nanoclusters. Calculated sizes using experimental band-edge shift with respect to that of the bulk value are also inscribed.

Fig 3. (Color online) Second order Raman scattering using UV (325 nm) excitation for ZnS nanoclusters with different cluster sizes. Inset shows the visible (532 nm) excitation for the corresponding samples. Spectra are shifted vertically for clarity.

Fig 4. (Color online) Multi-phonon spectra of ZnS nanoclusters of different cluster size using UV (325 nm) excitation showing overtones of optical modes. Spectra are shifted vertically for clarity. Inset shows the 2-TO modes for the 2.2 nm cluster corresponding to $A_1$ and $B_2$ symmetries under off-resonance conditions.

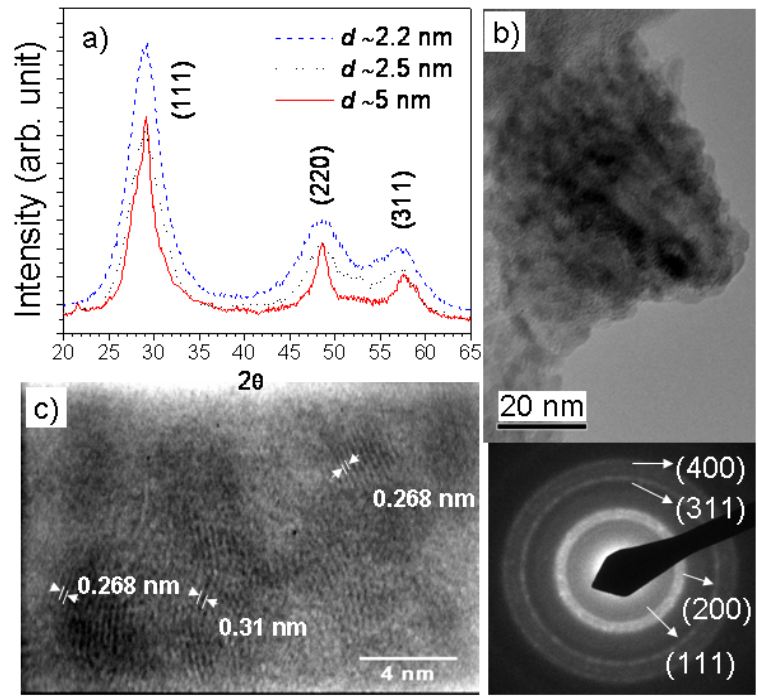

Fig. 1

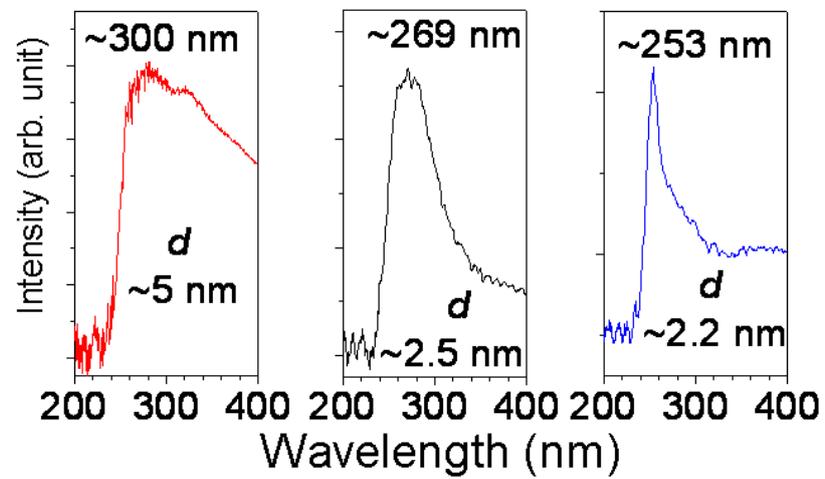

Fig. 2

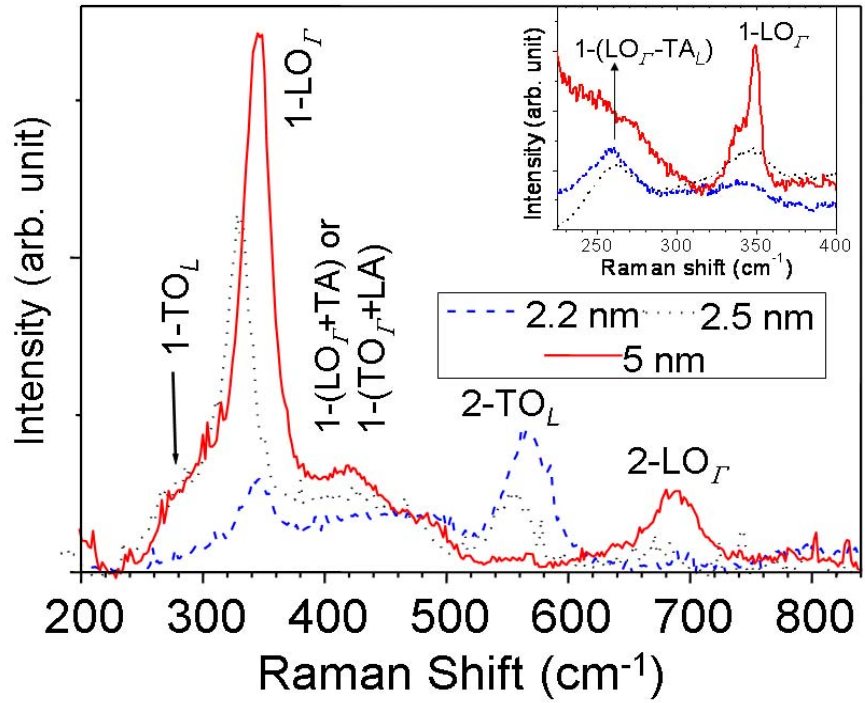

Fig. 3

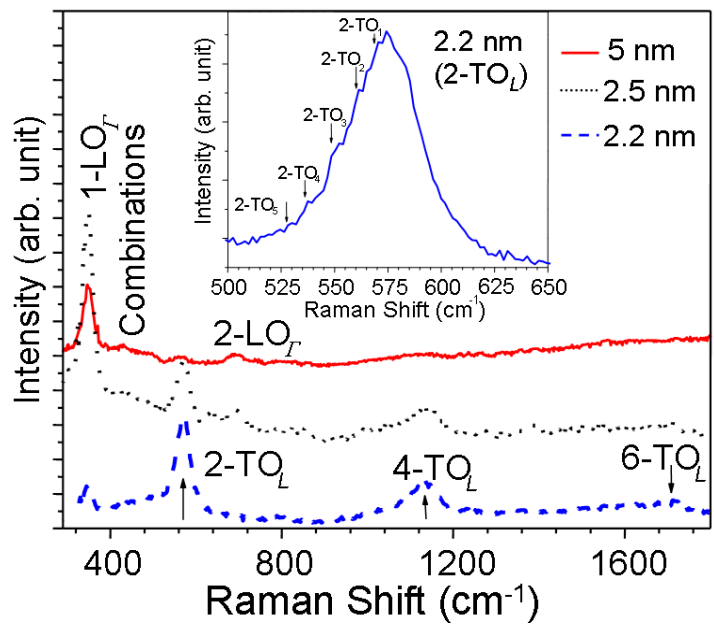

Fig. 4